# A Complementary Resistive Switch-based Crossbar Array Adder

A. Siemon, S. Menzel, *Member, IEEE,* R. Waser, *Member, IEEE* and E. Linn, *Member, IEEE*

*Abstract*—Redox-based resistive switching devices (ReRAM) are an emerging class of non-volatile storage elements suited for nanoscale memory applications. In terms of logic operations, ReRAM devices were suggested to be used as programmable interconnects, large-scale look-up tables or for sequential logic operations. However, without additional selector devices these approaches are not suited for use in large scale nanocrossbar memory arrays, which is the preferred architecture for ReRAM devices due to the minimum area consumption. To overcome this issue for the sequential logic approach, we recently introduced a novel concept, which is suited for passive crossbar arrays using complementary resistive switches (CRSs). CRS cells offer two high resistive storage states, and thus, parasitic 'sneak' currents are efficiently avoided. However, until now the CRS-based logic-in-memory approach was only shown to be able to perform basic Boolean logic operations using a single CRS cell. In this paper, we introduce two multi-bit adder schemes using the CRS-based logic-in-memory approach. We proof the concepts by means of SPICE simulations using a dynamical memristive device model of a ReRAM cell. Finally, we show the advantages of our novel adder concept in terms of step count and number of devices in comparison to a recently published adder approach, which applies the conventional ReRAM-based sequential logic concept introduced by Borghetti et al.

## I. INTRODUCTION

REDOX-BASED resistive switches (ReRAM) are considered as one of the most promising follower technologies for memory and logic applications [1]. In this technology the information is stored and calculated as two different non-volatile resistive states, a low resistive state (LRS) and a high resistive state (HRS). Two subclasses of ReRAM cells are most relevant for application. Whereas valence change mechanism (VCM) cells are based on oxygen vacancy movement in transition metal oxides (e.g. $TaO_x$ or $HfO_x$), electrochemical metallization (ECM) cells rely on the formation of a metallic Cu or Ag filaments [1, 2]. Both ECM and VCM cells offer a bipolar switching operation, i.e. SET and RESET occur at opposite voltage polarities. In 2008 Strukov et al. suggested to model ReRAM devices as memristive systems [3], sometimes also called memristor for short [4]. However, due to the complex physical mechanisms, memristive device modeling is challenging [5], and many available device models do not offer the required strong non-linear switching kinetics [6].

For memory applications a passive crossbar array is assumed to be the most favorable architecture, since it can offer a device area down to $4F^2$ [7]. However, due to absence of a transistor as selector device, low resistive devices in the matrix cause parasitic currents, also called current sneak paths, which drastically limits the maximum array size [8]. Thus, either a bipolar rectifying selector device or a complementary resistive switch (CRS) [9] configuration is required to enable passive arrays.

In terms of logic operations, there are three basic approaches based on ReRAM devices. The first one uses ReRAM devices as switchable interconnects. In the CMOL concept [10] for example, a sea of elementary CMOS cells, each consisting of two pass transistors and an inverter, is connected of discontinuous lines via ReRAM cells.

A second approach uses crossbar arrays for look-up-tables (LUT) for field programmable gate arrays (FPGA) applying small crossbar arrays. For example in [11] such architecture was suggested to implement a resistive programmable logic array (PLA) logic block realizing a full adder. Moreover, in [12, 13] a so-called memory-based computing approach using large crossbar arrays for multi-input-multi-output LUTs, which leads to reduced circuitry overhead, was suggested.

A completely different approach was suggested by Borghetti et al. [14] using ReRAM cells as conditionally switchable sequential logic devices, allowing logic-in-memory operations directly. This concept was further developed and adopted for CRS cells to improve array compatibility [15]. However, up to now only basic logic functions such as IMP or NAND have been shown for this approach by means of memristive simulations [16]. On the other hand, an adder concept using Borghetti's approach was suggested by Lehtonen et al. in [17]. Recently Kvatinsky et al. [18] represented two improved concepts. In this paper we show that advantageous adder concepts are feasible as well for our logic approach. These adder concepts are superior in terms of cycle and element count compared to the previous approaches. The paper is organized as follows: In section II the crossbar array

E. Linn, A. Siemon and R. Waser are with Institut für Werkstoffe der Elektrotechnik II (IWE II) & JARA-FIT, RWTH Aachen University, Sommerfeldstr. 24, 52074 Aachen, Germany (Corresponding Author e-mail: siemon@iwe.rwth-aachen.de)
S. Menzel and R. Waser are with Peter Grünberg Institut 7 (PGI-7) & JARA-FIT, Forschungszentrum Jülich GmbH, 52425 Jülich, Germany
The financial support of the German Research Foundation (DFG) under grant No. LI 2416/1-1 and SFB 917 is gratefully acknowledged.

nomenclature is introduced and the basic CRS logic concept is summarized. Then the inherent carry calculation capability of CRS devices is highlighted. In section III the novel adder schemes are explained, and in section IV the operation is verified by dynamical pulse simulations. In section V a comparison to Lehtonen's and Kvatinsky's adder approaches is drawn. Finally, in section VI the work is summarized and an outlook is given.

## II. COMPLEMENTARY RESISTIVE SWITCH-LOGIC

### A. Passive crossbar arrays

Ultra dense ReRAM-based memory architectures will be hybrid architectures with a standard CMOS component which is responsible for controlling the passive crossbar arrays. These arrays will be fabricated on top of the CMOS layers in the backend of line (BEOL) [7]. In general, the size of the crossbar arrays should be sufficiently large to justify the control circuit overhead. Thus, either appropriate selector devices are required at each cross point, or complementary resistive switches should be applied [9].

The basic idea underlying our approach is to extend the application of hybrid CMOS/crossbar architectures from pure memory operations towards logic-in-memory operations, by enabling a sequential access to the crossbar array devices [15]. Fig. 1a depicts a possible layout. The system could consist of many arrays and one control unit, which coordinates and addresses the signals to the specific wordlines (wl) and bitlines (bl). A typical array size could be for example 128 by 128 lines. Fig. 1a shows a system using CRS crossbar devices with only two arrays ($A_0$ and $A_1$) and an array size 3 by 5 to illustrate the basic concept. The structure of array A0 is depicted below this system section, showing that every intersection of a word- and bitline is a CRS cell. These CRS cells will be referred to as A$z$CRSwl$x$bl$y$ (cmp. Fig. 1), where A$z$ denotes the name of the array, in which the cell can be found, wl$x$ denotes the wordline of the cell and bl$y$ denotes the bitline. Thus the CRS cell A0CRSwl2bl0 is found in array $A_0$ at intersection $wl_2$ and $bl_0$.

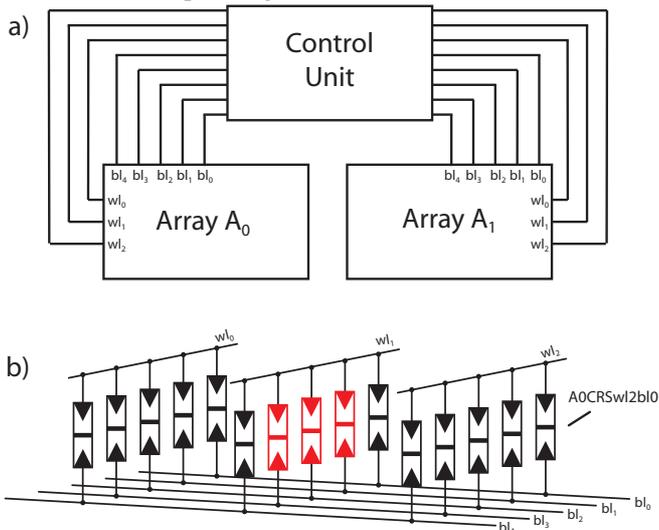

Fig. 1 (a) Expected system section layout, which consists of two Arrays ($A_0$ and $A_1$) and a control unit. (b) Each array has three wordlines ($wl_0$, $wl_1$ and $wl_2$) and five bitlines ($bl_0$, $bl_1$, $bl_2$, $bl_3$ and $bl_4$). The three red marked cells are used to compute a two bit addition.

The control unit enables free communication between all lines and is a key element for consecutive logic.

### B. Complementary Resistive Switches

CRS cells consist of two anti-serially connected ReRAM cells. A basic CRS operation in sweep mode is depicted in Fig. 2a. Both logic values '0' and '1' are represented by an in total high resistive state, since one cell is in HRS. '0' is represented by LRS/HRS and '1' by HRS/LRS. The 'ON' state is only a transition state, which is reached while changing the inner state from '0' to '1' or back. Here a half select scheme (e.g. [19]) is applied, so that there are three different voltage levels available at the word- and bitlines, low, high and ground. The devices need steep switching kinetics, since the devices must enable switching with the maximum voltage across the device for a given time period. Additionally, the cells must prevent switching if half of the maximum voltage is applied during the same time period. Note that a very steep switching kinetic is an intrinsic feature of resistive switching devices [20, 21], thus passive crossbar arrays are feasible.

### C. CRS single-bit logic operations

In [15] we introduced a CRS compatible 'stateful' logic approach. Fig. 2b represents a CRS cell as a finite state machine with two states. To switch from '0' to '1' the high potential, which is represented by the logical one '1', needs to be applied at the wordline and the low potential, logical zero '0', at the bitline of the cell. Otherwise the machine will stay in the '0'-state. To switch from '1' to '0' the low potential needs to be applied at the wordline and the high potential at the bitline of the cell. Otherwise the cell will stay in the '1'-state.

The general logic equation to represent this behavior is given by [15]:

$$Z = (wl \text{ RIMP } bl) Z' + (wl \text{ NIMP } bl) \overline{Z'} \qquad (1)$$

where $wl$ is the wordline connected to the device and $bl$ the bitline, $Z'$ is the device state prior to the application of the signals at wl and bl, and $Z$ is the device state after applying the signals. As follows, if the device is in state '1' ($Z'$ = '1'), the cell performs a reverse implication (RIMP) if the cell is in state '0' ($Z'$ = '0') an inverse implication (NIMP) is performed. 14 out of 16 Boolean functions are directly feasible within this approach [15]. The XOR and XNOR functions can only be realized with a second CRS cell. Note that a computation on more than one device is feasible, if the wl or bl input is the same for these computations on different devices.

Equation (1) must be considered as the basic equation to develop a synthesis tool for CRS-logic. For Borghetti's imply logic a few approaches for such a tool were presented [17, 22].

### D. CRS carry bit and sum bit calculation

An adder is the first step from basic logic operations towards complex arithmetic operations, since in CMOS all basic arithmetic operations (multiplier, divider and substractor) are in need of an adder. An adder consists of the possibility to calculate sum and carry bits. Fig. 2c depicts the truth tables of the carry and the sum function. In these functions the actual State $Z'$ is interpreted as the carry of significance i $c_i$, while the input variables $a_i$ and $b_i$ are the bits of the input words a and b with significance i. To compute $c_{i+1}$ $a_i$ and the negate of $b_i$ are applied to the wordline *wl* and bitline *bl*, respectively. Thus, using equation (1), the carry of the next higher significance $c_{i+1}$ can be calculated by the following equation in just one step:

$$c_{i+1} = \left(a_i \text{ RIMP } \overline{b_i}\right)c_i + \left(a_i \text{ NIMP } \overline{b_i}\right)\overline{c_i} \quad (2)$$

In the next few lines we show that this equation offers the correct result for $c_{i+1}$, which is in general expressed by:

$$c_{i+1} = a_i b_i + a_i c_i + b_i c_i \quad (3)$$

This can be rewritten as follows:

$$\begin{aligned}
c_{i+1} &= a_i b_i \left(c_i + \overline{c_i}\right) + a_i \left(b_i + \overline{b_i}\right) c_i + \left(a_i + \overline{a_i}\right) b_i c_i \\
&= \left(a_i + b_i\right) c_i + \left(a_i b_i\right) \overline{c_i} \\
&= \left(a_i \text{ RIMP } \overline{b_i}\right) c_i + \left(a_i \text{ NIMP } \overline{b_i}\right) \overline{c_i}
\end{aligned} \quad (4)$$

Thus, the carry calculation is an intrinsic feature of the CRS-logic.

In contrast, the sum needs two steps. First, actual state $Z'$ is interpreted again as the carry of significance i $c_i$. The input variables $a_i$ and $b_i$ are applied to the wordline *wl* and bitline *bl*, respectively, to calculate the intermediate state $s'_i$:

$$s'_i = \left(a_i \text{ RIMP } b_i\right) c_i + \left(a_i \text{ NIMP } b_i\right) \overline{c_i} \quad (5)$$

Next, $c_{i+1}$ is required as an input signal at the bitline, while $b_i$ is applied to the wordline:

$$s_i = \left(b_i \text{ RIMP } c_{i+1}\right) s'_i + \left(b_i \text{ NIMP } c_{i+1}\right) \overline{s'_i} \quad (6)$$

Note: It is favorable that the first sum computation step and the carry calculation step need the same input signal at the wordline, so both steps can be calculated at the same cycle in two different devices. Since the sum function needs $c_{i+1}$ as an input signal and only a destructive read-out is available, $c_{i+1}$ needs to be calculated in a different cell or needs to be written back.

The read-out scheme is depicted in Fig. 2d. A read-out is performed by applying '1' at the wl and '0' at the bl. Due to the fact that the state can be switched from '0' to '1' (destructive readout) it is possible that a write back step is needed. If a current spike is detected in the read-out cycle, the stored information is interpreted as a '0', if no current spike occurs the information is a '1'.

## III. ADDER SCHEMES

In this section we present two different bit-serial schemes to perform an addition on a CRS passive crossbar array by using simple consecutive signal sequences. By doing calculations in arrays instead of single cells, the main drawback of sequential logic, the need for multiple steps, can be eased, since array operations can be conducted in parallel.

Both adder schemes are based on the single-bit carry and sum calculation highlighted in section II.D. In this section, we introduce a way to perform multi-bit operations. Since CRS cells are passive devices there is no way, that they can pass information to the next stage. This is a major issue for complex calculations, which need more than one step or more than two input signals, like an adder. Hence either every intermediate step needs to be read out or the stored information is interpreted as a kind of 'third input' in the next step. As previously explained a read-out is destructive and requires a write back, if the data is needed later on. So the second possibility is preferable as it should be faster and more energy efficient. In fact, using parallel computing and stored information as a kind of 'third input' are the keys to designing a CRS adder.

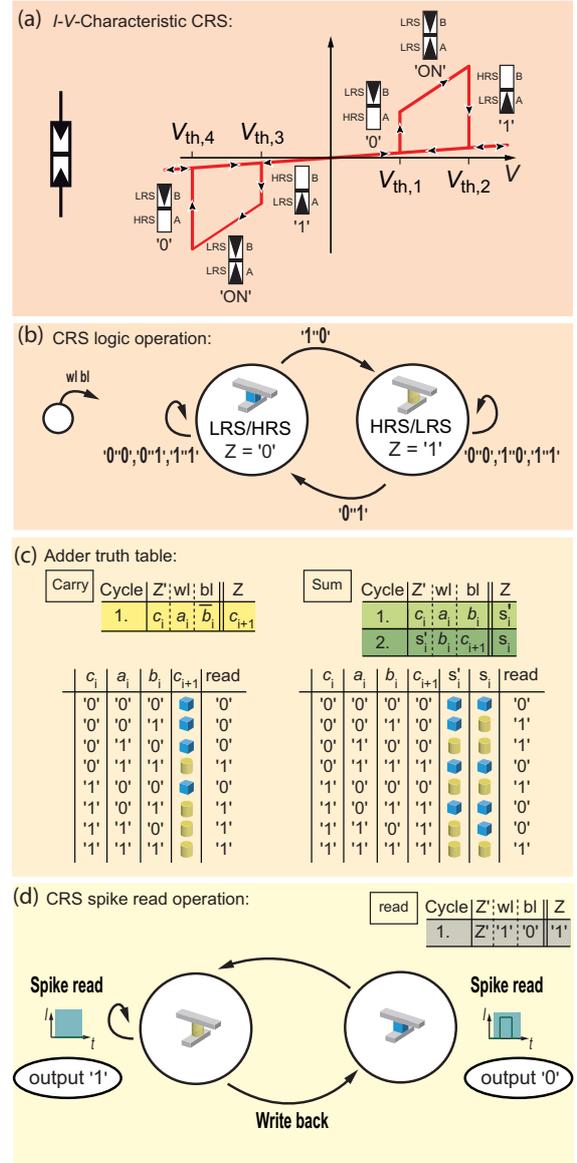

Fig. 2. (a) Basic CRS *I-V*-Characteristic. The logical state '0' is represented by the LRS/HRS state, logical '1' is represented by HRS/LRS and LRS/LRS is named 'ON-state' which is a transition state. The 'ON-window' is defined by $V_{th,2}$-$V_{th,1}$. (b) CRS as a finite state machine. The inputs at wordline wl and bitline bl are a high potential, represented by a logical one '1' and low potential represented by a logical zero '0'. (c) Truth tables for a carry and a sum functionality. The carry operation needs just one cycle (yellow), for which the actual state is interpreted as $c_i$ and the resulting state is $c_{i+1}$. The sum operation needs two cycles. In the first cycle (light green) the actual state is taken as $c_i$ and the resulting state is interpreted as the intermediate state $s'_i$. In the second step (dark green) the actual state is the previously calculated $s'_i$ and the resulting state is the sum bit $s_i$. Note that for the second step $c_{i+1}$ is needed as an input signal at the bitline, so $c_{i+1}$ needs to be calculated in another cell in a previous or in the same cycle. (d) Read-out operation (grey) for a CRS cell. A '0' was stored if a current spike (turquoise) is detected, if not it was a '1' (turquoise).

A difficulty in realizing an adder in CRS arrays was that there is no direct XOR-functionality available in CRS-logic [15]. But as shown before (cmp. Fig. 2c), it can be implemented in two steps by providing additional information from an auxiliary calculation, which is read out and used as an input signal.

Without loss of generality, we explain the schemes by means of a two bit addition. Since we operate a two's complement addition we need three devices to store the desired resulting word. For these examples we establish the following representation:

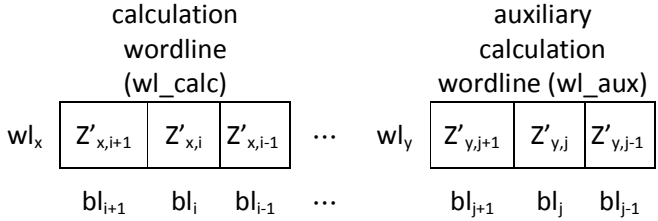

where $wl_x$ stands for the signal at the wordline wl with the number x in the calculation array, $bl_{i+1}$, $bl_i$ and $bl_{i-1}$ denote the signals at the bitlines with the numbers i+1, i and i-1 in the calculation array, $wl_y$ represent the signal at the wordline wl with the number y in the auxiliary calculation array. $bl_{j+1}$, $bl_j$ and $bl_{j-1}$ denote the signals at the bitlines with the numbers j+1, j and j-1 in the auxiliary calculation array and $Z'_{x,i+1}$, $Z'_{x,i}$, $Z'_{x,i-1}$, $Z'_{y,j+1}$, $Z'_{y,j}$ and $Z'_{y,j-1}$ denote the states prior to the application of the signals. This means, that the impact of the depicted signals is shown in the next step.

Without loss of generality we assume that the calculation takes place in the cells between wordline $wl_0$ and $bl_0$ to $bl_2$ or $bl_3$, respectively.

Note that not every cell is computing something in every cycle. If a cell should just keep the stored information until it is read out or further processed, the input signal at the bl is set to ground, which is represented by 0 due to the half select scheme.

*A. Precalculation-Adder*

This first approach needs two wordlines in two different arrays and requires the capability of reading and using an information bit in the same cycle. The sum is calculated in one wordline (wl_calc) the other wordline is used for auxiliary calculations (wl_aux). Without loss of generality wl_calc will be set to wordline $wl_0$ in array $A_0$ and wl_aux is set to wordline $wl_0$ in array $A_1$. These auxiliary calculations (precalculations) will be read out later in order to complete the computation of the final sum bits.

The needed operations can be grouped in three blocks: The initialization block (step 1-2), the preparation block (here step 3-5) and the finishing block (here step 6-8). In the initialization block, as the name states, the cells will be prepared to start the calculation. In the preparation block the cells prepare the final sum by calculating all needed information and intermediate states. In the final block the prepared information will be merged in the calculation wordline to finish the addition. The amount of steps of the second and third block depends on the input word length, while the first block is independent of it.

The operations of the precalculation-Adder (PC-Adder) in detail are:

**1. Step: Initialize/read-out**

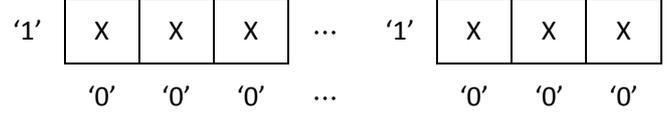

The first step is a read-out or initialization step during which the stored information is read out and the cells are brought to a known state '1'.

**2. Step: Programming $c_0$ in the calculation cells**

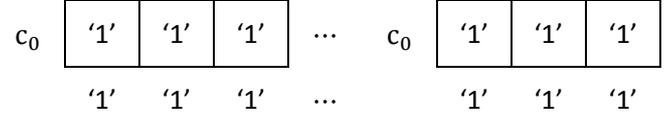

In the second step the first carry $c_0$ is programmed into all the calculation cells by setting the wordlines to $c_0$ and the bitlines to '1'. This step also enables distributed calculation and two's complement subtraction.

**3. Step: Calculation of $c_1$ and $s'_0$**

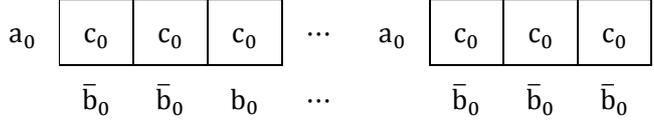

In the third step wl_calc calculates $c_1$ in all cells except for the least significant cell (A0CRSwl0bl0), which calculates the intermediate state $s'_0$ instead. This is done by setting the wl to $a_0$ and the bls to $\bar{b}_0$ or respectively $b_0$ (Fig. 2c). In wl_aux all cells calculate $c_1$, since this is the least significant carry needed to calculate the final sum bits. This is done by setting wl_aux to $a_0$ and the bls to $\bar{b}_0$. The least significant bit (LSB) cells (A0CRSwl0bl0 and A1CRSwl0bl0) are now ready for the last computational step and just store the current state until the auxiliary calculation is read out and the computational LSB is further processed.

**4. Step: Calculation of $c_2$ and $s'_1$**

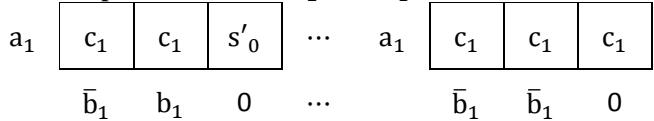

In the fourth step the most significant bit (MSB) cell (A0CRSwl0bl2) of wl_calc calculates $c_2$ and A0CRSwl0bl1 prepares the sum by calculating the intermediate state $s'_1$. This is nearly the same step as before but shifted one significance higher, so wl_calc is set to $a_1$, while $bl_2$ is set to $\bar{b}_1$ and $bl_1$ to $b_1$. In wl_aux the two cells of highest significance (A1CRSwl0bl2 and A1CRSwl0bl1) compute also $c_2$, by applying $a_1$ to wl_aux and $\bar{b}_1$ at $bl_2$ and $bl_1$.

**5. Step: Calculation of $c_3$ and $s'_2$**

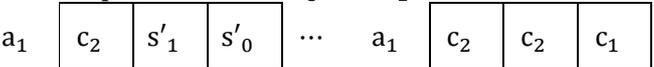

| $\overline{b_1}$ | 0 | 0 | ... | $\overline{b_1}$ | 0 | 0 |

In the last preparatory step in wl_calc only the MSB cell (A0CRSwl0bl2) calculates the intermediate state $s'_2$ by applying $a_1$ at the wordline and $b_1$ at $bl_2$. In wl_aux also only the MSB (A1CRSwl0bl2) needs to calculate $c_3$. This is done by applying $a_1$ once more at the wl_aux and $\overline{b_1}$ at $bl_2$. This step is necessary due to the doubled MSBs to secure a correct result.

**6. Step: Read-out auxiliary result $c_1$ and calculation of $s_0$**

| $b_0$ | $s'_2$ | $s'_1$ | $s'_0$ | ... | '1' | $c_3$ | $c_2$ | $c_1$ |
|       | 0 | 0 | $c_1$ | ... | 0 | 0 | '0' |

In the sixth step $s_0$ is calculated in wl_calc. For this the LSB of wl_aux is read out and is set as the input signal at $bl_0$ at wl_calc, while $b_0$ is applied at wl_calc.

**7. Step: Read-out auxiliary result $c_2$ and calculation of $s_1$**

| $b_1$ | $s'_2$ | $s'_1$ | $s_0$ | ... | '1' | $c_3$ | $c_2$ | '1' |
|       | 0 | $c_2$ | 0 | ... | 0 | '0' | 0 |

In the seventh and eighth step the same is done to calculate $s_1$ and $s_2$.

**8. Step: Read-out auxiliary result $c_3$ and calculation of $s_2$**

| $b_1$ | $s'_2$ | $s_1$ | $s_0$ | ... | '1' | $c_3$ | '1' | '1' |
|       | $c_3$ | 0 | 0 | ... | '0' | 0 | 0 |

After eight steps the sum is stored in wl_calc.
The result states are:

| $s_2$ | $s_1$ | $s_0$ | ... | '1' | '1' | '1' |

Depending on the bit length of the operands, the number of steps can be calculated as follows: $2(N+1)+2$, as can be seen from the cycle flow graph (Fig. 3).

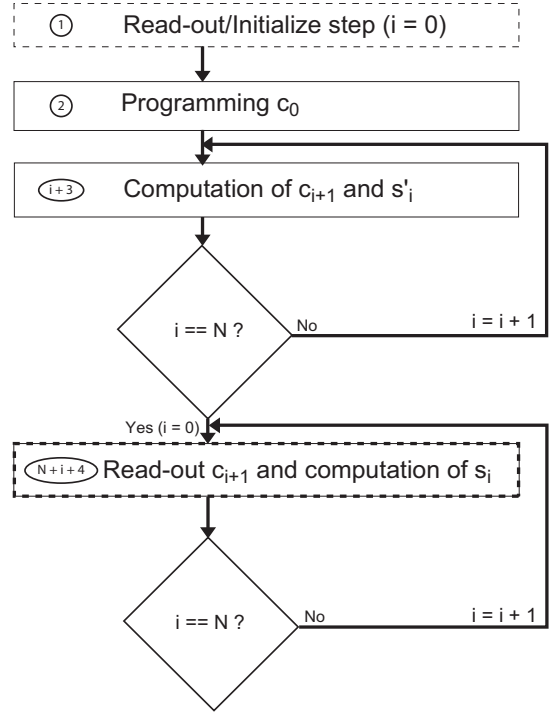

Fig. 3 Cycle flow graph of the Precalculation-Adder.

*B. Toggle-Cell-Adder*

In this paragraph we introduce an alternative implementation which only needs one wordline in one array, and so a fewer amount of cells. However, the number of required steps increases in this Toggle-Cell-Adder (TC-Adder) approach.

A difference to the first presented adder scheme is that not all cells in wl_calc will later be sum bits. In our presentation A0CRSwl0bl1 is the LSB cell. The A0CRSwl0bl0 cell is the toggle cell (TC), which calculates all carry bits and gives this scheme the name.

**1. Step: Initialize/read-out**

| '1' | X | X | X | X |
|     | '0' | '0' | '0' | '0' |

The first step is a read-out or initialization step, where the last information is read out and the cells are brought to a known state.

**2. Step: Programming $c_0$ in the calculation cells**

| $c_0$ | '1' | '1' | '1' | '1' |
|       | '1' | '1' | '1' | '1' |

In the second step the first carry $c_0$ is programmed in to all the calculation cells by setting wl to $c_0$ and the bls to '1'. This step enables distributed calculation and two's complement subtraction.

**3. Step: Calculation of $c_1$ and $s'_0$**

| $a_0$ | $c_0$ | $c_0$ | $c_0$ | $c_0$ |
|       | $\overline{b_0}$ | $\overline{b_0}$ | $b_0$ | $\overline{b_0}$ |

During the third step all cells except for the LSB cell calculate $c_1$ and the LSB cell is prepared for the sum bit by calculating the intermediate state $s'_0$. This is done by setting the wl to $a_0$ and the bls to $\bar{b}_0$ or $b_0$ respectively.

**4. Step: $c_1$ is read out**

| '1' | $c_1$ | $c_1$ | $s'_0$ | $c_1$ |
|---|---|---|---|---|
|  | 0 | 0 | 0 | '0' |

In the fourth step only the TC is read out.

**5. Step: Calculation of $s_0$**

| $b_0$ | $c_1$ | $c_1$ | $s'_0$ | '1' |
|---|---|---|---|---|
|  | 0 | 0 | $c_1$ | 0 |

In the fifth step $s_0$ is calculated in the LSB by applying $b_0$ at the wl and the read-out $c_1$ at $bl_1$.

**6. Step: Writing back $c_1$**

| $c_1$ | $c_1$ | $c_1$ | $s_0$ | '1' |
|---|---|---|---|---|
|  | 0 | 0 | 0 | '1' |

In the sixth step $c_1$ is written back to the TC.
Note that with this step the computation of the LSB is done and the information is just stored until it is read out.

**7. Step: Calculation of $c_2$ and $s'_1$**

| $a_1$ | $c_1$ | $c_1$ | $s_0$ | $c_1$ |
|---|---|---|---|---|
|  | $\bar{b}_1$ | $b_1$ | 0 | $\bar{b}_1$ |

In the seventh step the MSB cell (A0CRSwl0bl3) and TC calculate $c_2$, while A0CRSwl0bl2 computes $s'_1$, by applying $a_1$ at the wl and $\bar{b}_1$ and $b_1$ at the bls, respectively.

**8. Step: $c_2$ is read out**

| '1' | $c_2$ | $s'_1$ | $s_0$ | $c_2$ |
|---|---|---|---|---|
|  | 0 | 0 | 0 | '0' |

In the eighth step the TC is read out again.

**9. Step: Calculation of $s_1$**

| $b_1$ | $c_2$ | $s'_1$ | $s_0$ | '1' |
|---|---|---|---|---|
|  | 0 | $c_2$ | 0 | 0 |

In step nine $s_1$ is calculated in the A0CRSwl0bl2 cell by applying $b_1$ at the wl and the read-out $c_2$ at the $bl_2$.

**10. Step: Writing back $c_2$**

| $c_2$ | $c_2$ | $s_1$ | $s_0$ | '1' |
|---|---|---|---|---|
|  | 0 | 0 | 0 | '1' |

In the tenth step once again the TC is written back.

**11. Step: Calculation of $c_3$ and $s'_2$**

| $a_1$ | $c_2$ | $s_1$ | $s_0$ | $c_2$ |
|---|---|---|---|---|
|  | $b_1$ | 0 | 0 | $\bar{b}_1$ |

In the eleventh step the MSB is prepared by calculating the intermediate state $s'_2$. In the TC the last carry $c_3$ is computed. This step result out of the doubled MSBs to secure a correct result.

**12. Step: $c_3$ is read out**

| '1' | $s'_2$ | $s_1$ | $s_0$ | $c_3$ |
|---|---|---|---|---|
|  | 0 | 0 | 0 | '0' |

In step twelve the TC is read out the last time in this example.

**13. Step: Calculation of $s_2$**

| $b_1$ | $s'_2$ | $s_1$ | $s_0$ | '1' |
|---|---|---|---|---|
|  | $c_3$ | 0 | 0 | 0 |

In step thirteen the last sum bit $s_2$ is computed in the MSB by applying $b_1$ at wl and the read-out $c_3$ at $bl_3$.

After thirteen steps the sum is stored in the calculation cells, A0CRSwl0bl1, A0CRSwl0bl2 and A0CRSwlbl3.
The result states are:

| $s_2$ | $s_1$ | $s_0$ | '1' |
|---|---|---|---|

The cycle flow graph for the Toggle-Cell-Adder is slightly different compared to the PC-Adder, see Fig. 4. The amount of cycles increases to 4N+5 (PC-Adder: 2(*N*+1)+2), but only about the half of devices is required for this type of adder.

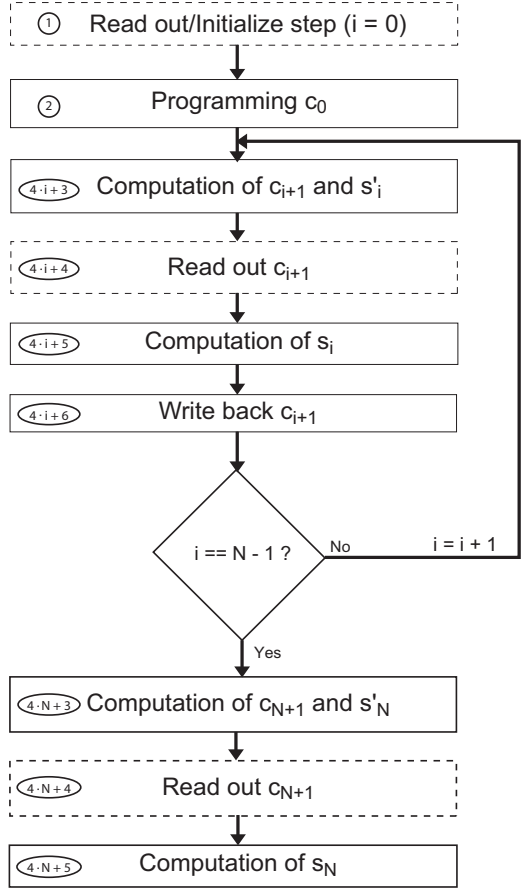

Fig. 4 Cycle flow graph of the Toggle-Cell-Adder.

## IV. ADDER SIMULATIONS

### A. ReRAM Device Modeling

An accurate, predictive and stable model is a key factor for future investigations concerning memory and logic designs. In [6] we defined three evaluation criteria, the *I-V* characteristic, the CRS *I-V* characteristic and the nonlinearity of the switching kinetics, and checked if different models fulfill these criteria. We showed that very few models could satisfactorily fulfill these criteria. So, accurate and predictive simulations, especially for VCM-type devices, are difficult to receive. However, for ECM devices there is a highly accurate memristive device model available [16, 23]. So the simulations are performed with this model to obtain a higher accuracy.

The switching mechanism of ECM devices is based on the electrochemically driven growths and dissolution in an ion conducting thin film. The electronic current is modulated by the variation of a tunneling gap between the filament tip and its counter electrode. In the ECM device model (cf. Fig. 5), a cylindrical Ag filament with a cross sectional area $A_{fil}$ is considered, which grows from the inert Pt towards the active Ag electrode within an insulating (switching) layer with thickness $L$. The dynamic evolution of the tunneling gap $x$ (the state variable) is driven by the ionic current $I_{ion}$ according to Faraday's law [16, 23]:

$$\frac{\partial x}{\partial t} = -\frac{M_{Me}}{zeA_{fil}\rho_{m,Me}} I_{ion}. \quad (7)$$

Here $M_{Me}$ is the molecular mass, $\rho_{m,Me}$ the mass density of the deposited metal, and $z$ the ionic charge of the cations.

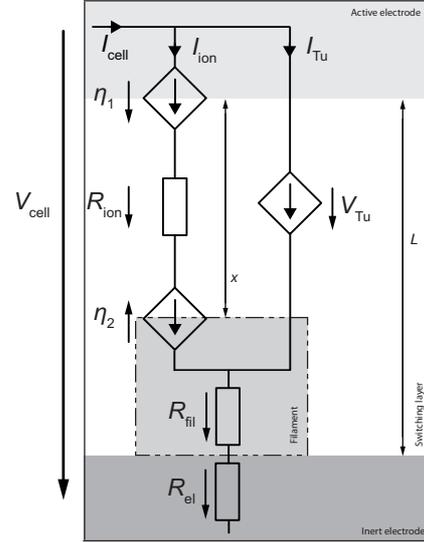

Fig. 5 Equivalent circuit model of the ECM cell.

For positive voltages $V_{cell}$ the gap $x$ decreases (SET) while it increases for negative currents (RESET). The ionic current path in the equivalent circuit model consists of two voltage controlled current sources $\eta_1$ and $\eta_2$, which resemble the oxidation/reduction reactions occurring at the active electrode/insulator and insulator/filament boundary, respectively. The ionic current $I_{ion}$ across the former interface is defined separately for positive and negative cell voltages according to the Tafel equation

$$I_{ion} = j_0 A_{fil} \begin{pmatrix} \exp\left(\frac{(1-\alpha)ze}{k_B T}\eta_1\right) - 1 & , \text{for } V_{cell} > 0 \\ 1 - \exp\left(-\frac{\alpha ze}{k_B T}\eta_1\right) & , \text{for } V_{cell} < 0 \end{pmatrix}. \quad (8)$$

Here $j_0$ is the exchange charge density, $\alpha$ is the charge transfer coefficient, and $\eta_1$ is the overpotential at the active electrode/insulator interface. For the insulator/filament interface the equations are defined with opposite polarities. Both, the ionic resistance $R_{ion} = x/(\sigma_{ion}A_{fil})$, which models the ion drift within the insulator, and the filament resistance $R_{fil} = (L-x)/(\sigma_{fil}A_{fil})$ are assumed to be ohmic. Note that the electronic current is controlled by the gap size $x$, and is defined as a tunneling current:

$$I_{Tu} = \frac{3\sqrt{2m_{eff}\Delta W_0}}{2x}\left(\frac{e}{h}\right)^2 \exp\left(-\frac{4\pi x}{h}\sqrt{2m_{eff}\Delta W_0}\right) A_{fil} V_{Tu}, \quad (9)$$

where $m_{eff} = m_r m_0$ is the effective electron tunneling mass, $\Delta W_0$ the barrier height and $h$ the Planck's constant. Using the equivalent circuit diagram the equation system (7) – (9) is implemented in VerilogA and solved by using Spectre. Table 1 summarizes the simulation parameters. Here, we used the same parameters as in our previous model [16] apart from a larger filament area.

| Parameter | Symbol | Value |
|---|---|---|
| Resistance of the electrodes | $R_{el}$ | 70 mΩ |
| Switching layer thickness | $L$ | 20 nm |
| Mass density | $\rho_{m,Me}$ | 8.95 g cm$^{-3}$ |
| Filament area | $A_{fil}$ | 135.87 nm$^2$ |
| Molecular mass | $M_{Me}$ | 1.06 x 10$^{-22}$ g |
| Conductivity of the active electrode / filament | $\sigma_{fil}$ | 5 x 10$^7$ S m$^{-1}$ |
| Ionic conductivity of the switching layer | $\sigma_{ion}$ | 1 x 10$^2$ S m$^{-1}$ |
| Barrier height | $\Delta W_0$ | 3.6 eV |
| Effective electron tunneling mass | $m_{eff}$ | 0.86 x 9.1 x 10$^{-31}$ kg |
| Temperature | $T$ | 300 K |
| Charge transfer coefficient | $\alpha$ | 0.5 |
| Ionic charge of the cations | $z$ | 1 |
| Exchange current density | $j_0$ | 0.01 A m$^{-2}$ |

Table 1 Simulation parameters

In Fig. 6 a simulated unit cell *I-V* characteristic and a CRS *I-V* characteristic are depicted. A typical feature of an ECM device is the asymmetry of SET (~1.3 V) and RESET (~-0.5 V) voltages (Fig. 6a). This asymmetry leads to a reduced ON-window width for the CRS device (Fig. 6b). However, the actual width of the ON-window has no impact on the device performance since the LRS/LRS state is not accessed neither for write nor read: in the applied Spike read scheme (Fig. 2d) a full switching from LRS/HRS to HRS/LRS is performed. Note that a read scheme using the LRS/LRS state ('Level read') would be also feasible due to the nonlinearity of the device kinetics, see [16, 24]. Furthermore, in [17] we showed that the ON-window can be adjusted by adding a serial resistance.

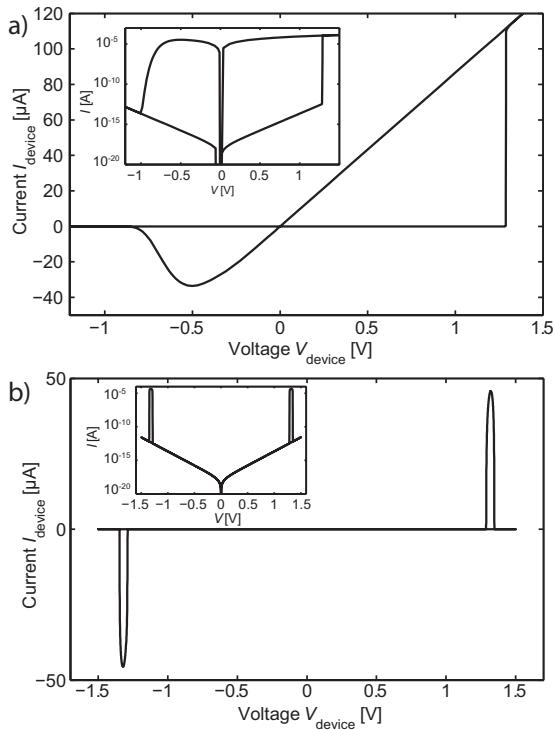

Fig. 6 (a) Simulated *I-V* characteristic of an ECM ReRAM device. (b) Simulated *I-V* characteristic of a corresponding CRS cell. Insets show characteristics on a log scale.

Based on this dynamic ECM model, VerilogA simulations of a crossbar array implementing the adder schemes introduced above were conducted to prove the basic concept. The control unit functionality was done manually by adjusting the input signals.

*B. Simulation of the Precalculation-Adder*

In Fig. 7 the simulation results of the Precalculation-Adder scheme are depicted. It shows the simulation for an exemplary addition. In this example the inputs are set to a = 01 and b = 01. The first line of both arrays depicts the potential at the calculating wordline. The second, fourth and sixth lines show the potential at the calculating bitlines and the third, fifth and seventh line show the current corresponding to the bitlines. Here the array $A_0$ is the calculating array, while array $A_1$ calculates the auxiliary calculations. The calculation wordlines are set in both arrays to wordline $wl_1$, while the calculating bitlines are set to $bl_1$, $bl_2$ and $bl_3$. The background colors of the steps are correspondingly set to Fig. 2c, so the first step (grey) performs a read-out on all cells. The second step (orange) programs $c_0$ in all calculation cells. In the third step array $A_0$ calculates $c_1$ (yellow) and $s'_0$ (light green) and $A_1$ calculates $c_1$ (yellow) in all cells. In the fourth step A0wl1bl3, A1wl1bl3 and A1wl1bl2 compute $c_2$ (yellow) and A0wl1bl2 computes $s'_1$ (light green). In the fifth step only $s'_2$ (light green) is computed in array $A_0$ and $c_3$ (yellow) is calculated in array $A_1$. After these preparatory steps the in array $A_1$ calculated information is read out and used as input signals at the bls of array $A_0$. So the sixth step presents the read-out (grey) of the information stored in A1CRSwl1bl1 and the calculation of the first sum bit $s_0$ (dark green). The information was interpreted as a one '1' (turquoise), since no current spike occurred. In the seventh step the next stored information is read out (grey). Here a current spike occurs (turquoise), thus the information is interpreted as zero '0' and used as the input signal at $bl_2$ in array $A_0$ in order to calculate the sum bit $s_1$ (dark green). In the last computational step the next information is read out (grey) again. A current spike (turquoise) occurs so the information is interpreted as zero '0' and set to an input signal at $bl_3$ to calculate the last sum bit $s_2$ (dark green). Then another read-out step (grey) is performed to show that the stored information is the desired result.

*C. Simulation of the Toggle-Cell-Adder*

In Fig. 8 the simulation results of the Toggle-Cell-Adder scheme are depicted. It shows the simulation for an exemplary addition. In this example the inputs are set to a = 01 and b = 01. The first line depicts the potential at the calculating wordline, here $wl_1$ in Array $A_0$. The second, fourth and sixth lines show the potential at the calculating bitlines, $bl_4$, $bl_3$ and $bl_2$. The eighth line depict the potential at the TC bitline. The third, fifth, seventh and ninth line show the current corresponding to the bitlines. The background colors of the steps are correspondingly set to Fig. 2c, so the first step (grey) performs a read-out on all cells. The second step (orange) programs $c_0$ in all calculation cells and in the TC. In the third step $c_1$ (yellow) and $s'_0$ (light green) are being calculated. In the fourth step a read-out (grey) of the TC is performed. No current spike (turquoise) occurs so the stored information is interpreted as a one.

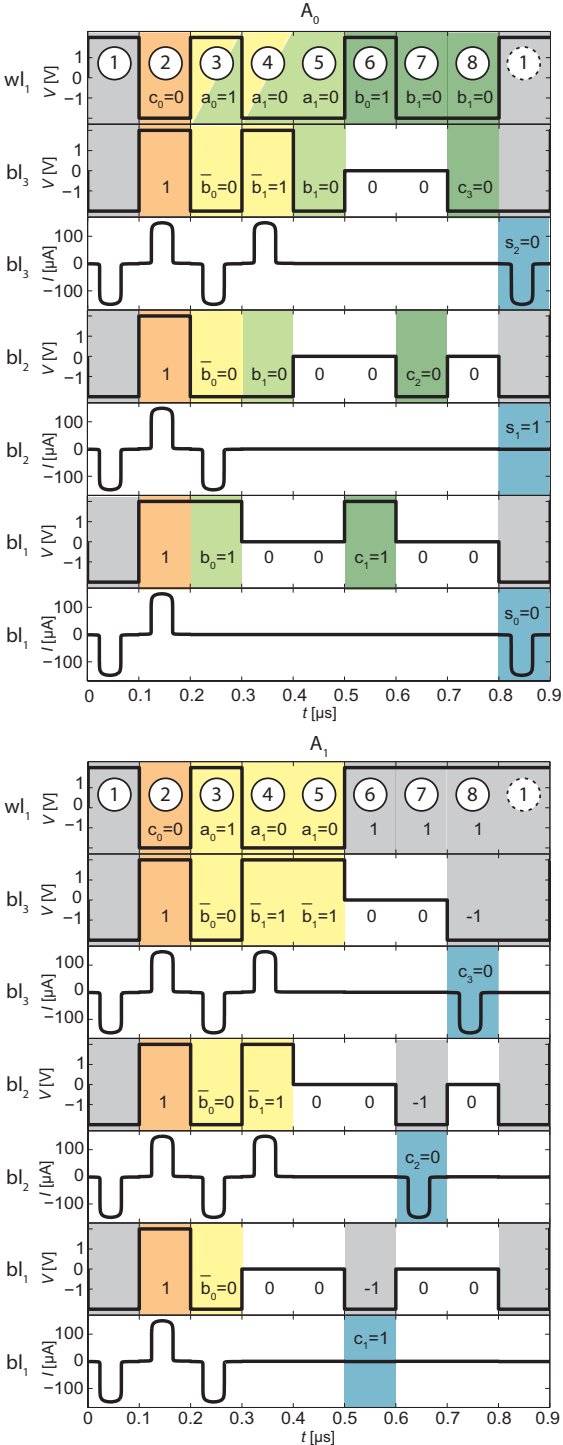

Fig. 7 Simulation of the Precalculation-Adder with inputs a = 01 and b = 01. The background color is set correspondingly to Fig. 2. Yellow depicts the carry functionality, light green the first sum computation cycle, dark green the second step of the sum function, grey a read-out step, turquoise the read-out current response and orange the programming step. Steps 1-2 are initialization steps, after these $c_0$ is programmed in every cell. Steps 3-5 are the preparatory steps in which all needed information, $c_{i+1}$ (yellow) and $s'_i$ (light green), are calculated. In steps 6-8 the information is merged and the final sum bits (dark green) are calculated. The last step is another read-out to show that the stored information is the desired result.

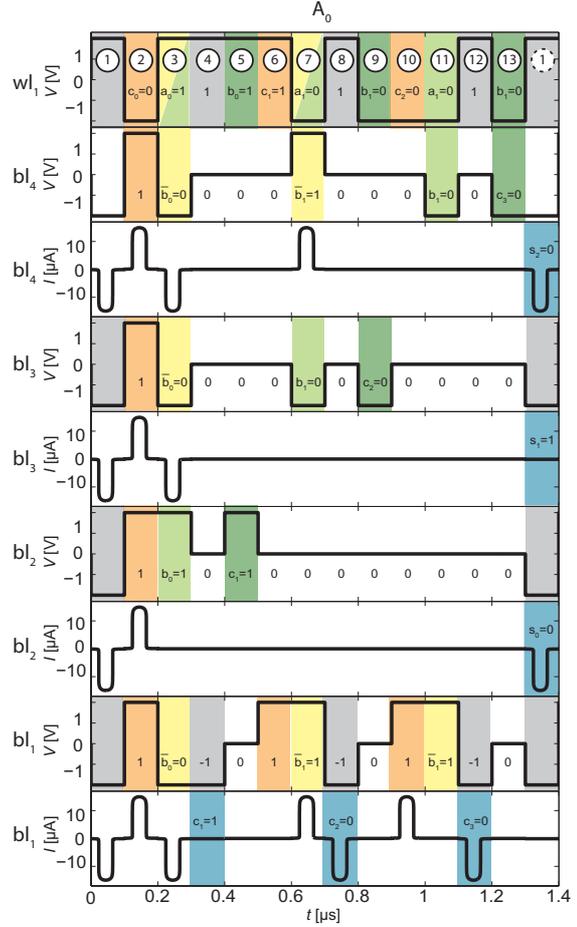

Fig. 8 Simulation of the Toggle-Cell-Adder with inputs a = 01 and b = 01. The background color is set correspondingly to Fig. 2. Yellow depicts the carry functionality, light green the first sum computation cycle, dark green the second step of the sum function, grey a read-out step, turquoise the read out current response and orange the programming step. Steps 1-2 are initialization steps, after these $c_0$ is programmed in every cell. In Steps 3, 7 and 11 the needed information, $c_{i+1}$ (yellow) and $s'_i$ (light green), are calculated. In steps 4, 8 and 12 the TC is read out (grey). In steps 5, 9 and 13 the read out information is used to calculate the final sum bits (dark green). In steps 6 and 10 the read out information is written back in the TC (orange) to enable further calculation. The last step is another read out step to show that the stored information is the desired result.

The read-out information is used in the fifth step to compute the final sum bit $s_0$ (dark green). In step six the read-out information is written back (orange) in the TC. In the seventh step $c_2$ (yellow) and $s'_1$ (light green) is computed. Once more the TC is read out (grey) in the eighth step. The stored information is interpreted as a zero, since a current spike (turquoise) is detected. This read-out information is used to calculate the final sum bit $s_1$ (dark green) in the ninth step. In the tenth step the TC is programmed back with the read-out information (orange). In the eleventh step $c_3$ (yellow) and $s'_2$ (light green) are computed. In step twelve the TC is read out (grey) for the last time in this example. The occurring current spike (turquoise) is interpreted as a zero. In step thirteen this readout information is used to calculate the last final sum bit $s_2$ (dark green). At last another read-out step (grey) is performed to show that the stored information is the desired result.

## V. Adder Comparison

In [17] Lehtonen et al. introduced an adder concept using the imply logic approach according to Borghetti [14]. In [18] Kvatinsky et al. introduced two new, improved adder designs, a parallel and a serial. First, we want to compare these three approaches with our adder schemes in terms of cycles and device count (see Table 2). Moreover, a third criteria referring to the compatibility with common crossbar arrays is also considered in Table 2.

One can easily see that the newly introduced schemes require fewer devices and cycles. Also a very important fact is that these adder work on $4F^2$ passive crossbar arrays. In contrast, Kvatinsky's parallel approach needs a more complex crossbar architecture.

|  | Lehtonen [17] | Kvatinsky [18] | | New approaches in this work | |
|---|---|---|---|---|---|
|  |  | Serial | Parallel | PC-Adder | TC-Adder |
| No. devices | 3N+5 | 3N+3 | 9N | 2(N+1) | N+2 |
| No. Cycles | 88N+48 | 29N | 5N+18 | 2(N+1)+2 | 4N+5 |
| Common Crossbar | Yes | Yes | No | Yes | Yes |

Table 2 Comparison between [17], [18] and the new approaches of this work. Best values in each line are marked by green background color, intermediate values by grey color and worst values by red color.

Furthermore, we want to compare the introduced schemes in requirements of the interconnect. The Precalculation-Adder needs a global interconnect, so that the read-out information from one array can be applied as an input in another array. Since we assume that we can read out one cell and apply this information at another bitline during one cycle the cycle duration and the control unit functionality needs to be adjusted to this case.

The TC- Adder does not necessarily need a global interconnect, since the read-out information was stored in the same wordline. Nevertheless either if no near interconnect is present the information has also to be sent to the control unit or if a near interconnect is present the information needs to be stored in a register, since it will be needed in two more cycles. But even if the information is sent to the control unit the cycle duration is not very crucial, since the information is needed just in the next cycle.

## VI. Conclusion

In this work we have presented two novel CRS-based adder schemes which enable arithmetic operations within passive crossbar memories. The Toogle-Cell-Adder offers the lowest amount of cells while the Precalculation-Adder requires the fewest number of steps. Both concepts enable multi-bit logic operations in CRS arrays in a very efficient manner and could pave the path to new computing architectures based on ReRAM-type nanocrossbar memories.